\documentclass[12pt,preprint]{aastex}
\usepackage{natbib}
\begin{document}

\title{Time-Variable Accretion in the TW Hya Star/Disk System}

\author{J. A.  Eisner\altaffilmark{1}, G. W. Doppmann\altaffilmark{2},
J. R. Najita\altaffilmark{2}, D. McCarthy\altaffilmark{1}, C. Kulesa\altaffilmark{1}, 
B. J. Swift\altaffilmark{1}, \& J. Teske\altaffilmark{1}}

\altaffiltext{1}{Steward Observatory,
University of Arizona, 933 N Cherry Avenue, Tucson, AZ 85721}
\altaffiltext{2}{National Optical Astronomy Observatory, 950 N. Cherry Avenue,
Tucson, AZ 86719}
\email{jeisner@email.arizona.edu}

\keywords{stars: planetary systems---stars: pre-main-sequence---stars:
  individual (TW Hya)---techniques:spectroscopic}

\begin{abstract}
We present two epochs of observations of TW Hya from the high-dispersion near-IR 
spectrograph ARIES at the MMT.  We detect strong emission from the Br$\gamma$
transition of hydrogen, indicating an accretion rate substantially
larger than previously estimated using hydrogen line emission.
The Br$\gamma$ line-strength varies across our two observed epochs.
We also measure circumstellar-to-stellar flux ratios (i.e., veilings)
that appear close to zero in both epochs.  
These findings suggest that TW Hya experiences episodes of enhanced
accretion while the inner disk remains largely devoid of dust.  We
discuss several physical mechanisms that may explain these
observations.
\end{abstract}

\section{Introduction \label{sec:intro}}
TW Hya is a nearby ($\sim 56$ pc), young 
($\sim 10$ Myr) star \citep{WEBB+99} 
surrounded by an accretion disk that evinces a large inner hole
as judged from the observed spectral energy distribution 
\citep[SED;][]{CALVET+02} and mid-IR and sub-mm imaging
\citep{RATZKA+07,HUGHES+07}.   These data can be modeled with an
optically thick disk whose inner edge is located $\sim 4$ AU from the central star.  
While observations of 10 $\mu$m silicate emission \citep{SLR00,UCHIDA+04}
together with spatially-resolved emission at 2 $\mu$m \citep{ECH06} suggest 
the presence of dust grains with sizes less than a few microns at 
stellocentric distances $R \la 4$ AU, this inner disk material appears 
optically thin, and has been estimated to constitute less than a lunar mass 
\citep{CALVET+02}.  Such a small amount of material is compatible with
the very low near-IR veiling measured previously
\citep{JV01,YJV05}.

Warm gas has also been detected within the optically thin inner region
of TW Hya's circumstellar disk
\citep{HERCZEG+04,RETTIG+04,SALYK+07,SALYK+09,PONTOPPIDAN+08,NAJITA+06}.
In fact, the gas appears to be distributed 
in a disk with a similar inclination to the optically thick dust disk observed
at larger radii \citep[$i \sim
3$--7$^{\circ}$;][]{PONTOPPIDAN+08,QI+04}.  The inner disk of TW Hya,
while optically thin, is clearly not completely devoid of matter.

A variety of accretion signatures have been detected in TW Hya, indicating 
some flow of gaseous material through the inner disk.  Observations of
H$\alpha$ emission, interpreted in the context of a magnetospheric accretion
model, indicated $\dot{M} = 5 \times 10^{-10}$ M$_{\odot}$ yr$^{-1}$ 
\citep{MUZEROLLE+00}.  A shock model fitted to UV and optical photometry
indicated a consistent value of $\dot{M}=4 \times 10^{-10}$  M$_{\odot}$ 
yr$^{-1}$ \citep{MUZEROLLE+00}.   A similar model fitted to a separate
UV spectrophotometric dataset yielded  $\dot{M}=2 \times 10^{-9}$  M$_{\odot}$ 
yr$^{-1}$ \citep{HERCZEG+04}.  Analysis of two additional epochs of UV
data separated by approximately two years indicate an accretion rate of 
$\sim 1.5 \times 10^{-9}$ M$_{\odot}$ yr$^{-1}$ on both occasions \citep{HH08}. 
Still higher accretion rates were suggested by modeling of X-ray emission from 
TW Hya:  $\dot{M} \sim 10^{-8}$ M$_{\odot}$ yr$^{-1}$ \citep{KW02}.
In contrast, a subsequent study of X-ray emission derived
an accretion rate of  $\sim 10^{-11}$  M$_{\odot}$ yr$^{-1}$ \citep{SS04}.
Modeling of the C$_{\rm IV}$ $\lambda 1549$ \AA $\:$ line flux yields
$\dot{M}=4 \pm 2 \times 10^{-8}$  M$_{\odot}$ yr$^{-1}$ \citep{VJL00}.

%If one assumes a viscous disk with $\alpha=0.01$ and a gas-to-dust ratio
%of 100, then the measured
%dust surface density \citep{WILNER+00,WILNER+05} can be converted
%into an accretion rate $\ga 10^{-8}$ M$_{\odot}$ yr$^{-1}$ \citep[although
%there is reason to suspect a much lower value of $\alpha$ for TW Hya; e.g.,][]{ECH06}.

While the large dispersion in accretion rate measurements is due in large part 
to different systematic effects for different techniques, in some cases
the same technique used on different epochs (generally by different investigators)
has yielded different accretion rates.  Moreover, a study of Na D line variability 
suggested accretion rates varying between 
$\sim 10^{-9}$ and $10^{-8}$  M$_{\odot}$ yr$^{-1}$ over a 1-yr period
\citep{AB02}.  Thus, it seems that the amount of material in the inner
disk may fluctuate significantly over time.

In this paper, we present observations of Br$\gamma$
emission and near-IR veiling. The two epochs of data show 
significant variations in Br$\gamma$ emission strength, but not in veiling.
We use these findings to examine the viability to
different mechanisms proposed to maintain the optically thin inner
region of TW Hya's disk.

\section{Observations and Data Reduction \label{sec:obs}}
We observed TW Hya on UT 17 January 2009 and UT 8 May 2009 with the ARIES
instrument at the MMT.  ARIES receives a corrected beam from the
adaptive secondary of the MMT, and performs imaging and spectroscopy
from 1 to 5 $\mu$m wavelengths (although the 3--5 $\mu$m detector is not
yet implemented).   We used the echelle mode of ARIES, with a 
resolving power of $R=30,000$.
The slit is $1'' \times 0\rlap{.}''1$,
well-matched to the diffraction-limited, AO-corrected resolution
of the MMT.   The small slit allows many orders to be placed
on the detector simultaneously, and our setup included $>25$
non-contiguous orders
across the $H$ and $K$ bands.

Based on the echelle dispersion, we expect a velocity resolution
of approximately 10 km s$^{-1}$.  Measured FWHMs of argon lamp lines
show that the actual instrumental velocity resolution of these observations
is approximately 14 km s$^{-1}$.

TW Hya was observed for 10 minutes on 17 January 2009 and for 40
minutes on 8 May 2009 (both of these are on-source integration times).  
In each epoch, observations of TW Hya 
were interleaved with observations, at  similar airmass, 
of an A1V telluric calibrator star, HIP 54682.
The data were flat-fielded using spectra of an incandescent lamp, and
extracted, wavelength calibrated, and telluric corrected, using IRAF
packages \citep{MVB92}.
Wavelength calibration used telluric absorption lines and the HITRAN
linelist \citep{ROTHMAN+05}, and is
accurate to 1--2 km s$^{-1}$.

We also obtained imaging data of TW Hya and the nearby star TWA 7 during 
both epochs.  TWA 7 is a weak-line T Tauri star \citep{WEBB+99} of
similar age to TW Hya \citep{NEUHAUSER+00}.  It
exhibits no detectable emission above the stellar
photosphere in the near-IR, although it does have long-wavelength
excess emission indicative of a debris disk \citep{MKW07}.
Due to the low declinations of both TW Hya and TWA 7, we did not have
time to obtain spectra of TWA 7.  However, since both
sources were observed at similar airmasses and have similar colors,
TWA 7 can be used for relative photometric calibration.
In January, 2009, both objects
were observed using the $K_{\rm s}$ filter with integration times of 0.9 s.
In May, 2009, the atmospheric transparency was better, and these objects
saturated the detector when the $K_{\rm s}$ filter was used, even for the shortest
integration times.  We therefore observed TW Hya and TWA 7
with a narrowband H$_2$ filter, using 5 s integrations, during the May observations.

During our May observations, we also obtained spectra of GJ 568 A, an M3
V star \citep{GRAY+03}.  The total integration time was 25 minutes.
The spectra were calibrated using the same procedure described above
for TW Hya.  We use GJ 568 below for a rough comparison of the veiling
observed in TW Hya with that of an unveiled main-sequence star.

\section{Results and Analysis \label{sec:results}}

\subsection{Photometry \label{sec:phot}}
For each observed epoch, we computed the counts for both TW Hya
and TWA 7 within an aperture of 40-pixel  ($1\rlap{.}''6$) radius.
The sky value was determined as the median of pixels with radii
between 60 and 80; this value was subtracted from each pixel in
the photometric aperture.  Assuming the 2MASS magnitude of TWA 7, 
$m_K=6.90$, we derive the $K$ magnitude of TW Hya at both
epochs (Table \ref{tab:results}).  The two measurements are
consistent with each other, and with the 2MASS magnitude, $m_K=7.30$.

\subsection{Br$\gamma$ Spectra}
\begin{figure}[tbhp]
\plotone{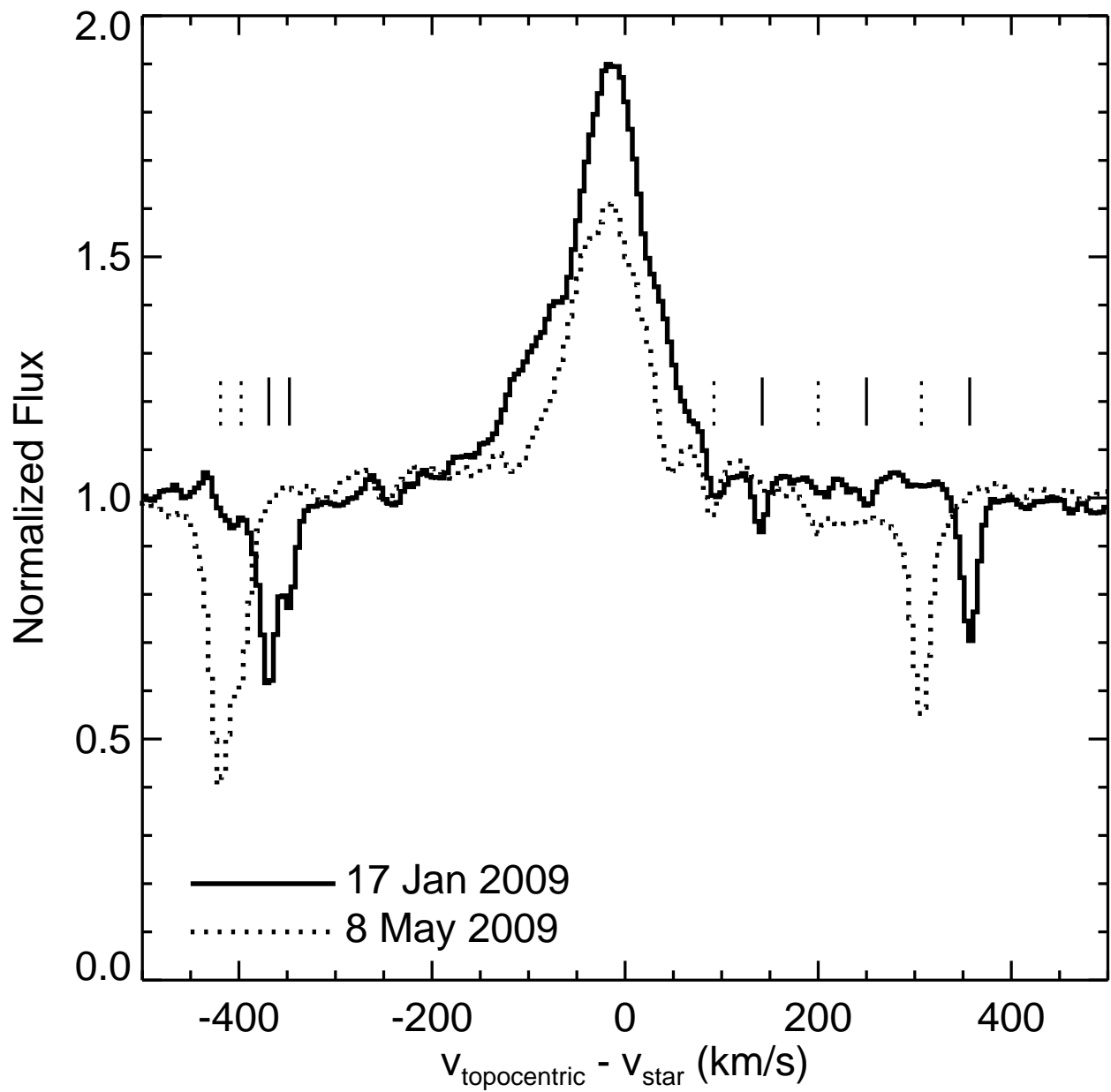}
\caption{Spectra of Br$\gamma$ emission from TW Hya observed with
ARIES at the MMT on 17 January 2009 (solid histogram) and 8 May 2009
(dotted histogram).  Telluric absorption features have not been
divided out from these spectra, and the velocities of telluric lines
in both epochs are indicated by vertical lines.  The stellar velocities used to create
the velocity axes in these plots were determined by cross-correlating
observed spectra of the CO bandheads with model spectra, and are accurate to
$\sim 5$ km s$^{-1}$.
\label{fig:brg}}
\end{figure}

Our Br$\gamma$ spectra are shown in Figure \ref{fig:brg}.  We measure
equivalent widths (EWs) from these spectra by integrating the line
flux between 2.1644 and 2.1668 $\mu$m; this range covers all of the
Br$\gamma$ emission seen in both epochs.  The main uncertainty in our
EW determination is the continuum level.  We estimate this
uncertainty by fitting the continuum in different spectral regions.
We find 12\% error bars for our EW measurements.

The lines measured in both epochs are blue-shifted.  In the January data,
the blue-shift is more pronounced, and a ``blue shoulder'' is more
prominent in the spectrum.  Blue-shifts and blue-shoulder-profiles are
expected for models of infalling gas, since absorption of Br$\gamma$
emission by the infalling gas along the line of sight preferentially removes red
emission \citep[e.g.,][]{HHC94,NCT96,MHC98}.  

The emission in both
epochs is substantially broader and more asymmetric than expected from
stellar flares, which typically show strong (H$\alpha$ line-to-continuum ratios
$\ga 10$), narrow (FWHM $<50$ km s$^{-1}$), symmetric emission line cores 
\citep[e.g.,][]{HP91,HILTON+10}.  Similarly, quiescent chromospheric
activity produces emission that is narrower and more symmetric than
emission produced by accretion processes \citep{WB03}. 
Since the difference spectrum for
the two epochs is broad and asymmetric--lacking a strong, narrow core--the change
between spectra is probably not due to chromospheric activity. 

Using the broad-band magnitude of TW Hya derived in \S \ref{sec:phot},
we can convert the observed Br$\gamma$ EWs into line luminosities.
These, in turn, can be converted into accretion luminosities using
an empirical relationship derived by \citep{MHC98}.  To convert the accretion
luminosities into mass accretion rates requires knowledge of the stellar
parameters: $\dot{M} \approx L_{\rm acc} R_{\ast} / GM_{\ast}$.  For this
calculation, we use the stellar parameters derived by \citet{WEBB+99},
assuming a distance of 56 pc:
$M_{\ast}=0.7$ M$_{\odot}$, $R_{\ast}=1$ R$_{\odot}$, and $T_{\rm eff}=4000$ K.
These are the same values used by \citet{MUZEROLLE+00}, enabling
a fair comparison with their results.  Derived line luminosities, 
accretion luminosities, and accretion rates are listed in Table \ref{tab:results}.

\subsection{Veiling \label{sec:veiling}}
We estimate continuum veiling in the $K$-band using Mg and Al lines near 2.11
$\mu$m.    These lines are found in a
spectral region relatively free of telluric lines (unlike, for
instance, the CO ro-vibrational bandheads), and have been used
in past studies for veiling measurements of young stars
\citep[e.g.,][]{DOPPMANN+05,EISNER+07c}.

The basic method we use to estimate veiling is to compare the observed
spectrum to a synthetic spectrum for a stellar photospheric model.
We first determine oscillator strengths for the Mg and Al transitions 
by comparing synthetic Nextgen spectra for a solar-type star to
observed solar spectra.  Synthetic spectra relevant for TW Hya are
then computed using stellar parameters determined from previous high
dispersion optical spectroscopy \citep{YJV05}: $T_{\rm eff} = 4126 \pm
24$ K, $\log g = 4.84 \pm 0.16$, [M/H]$= -0.11 \pm 0.13$, and $v \sin
i = 5.80 \pm 0.63$ km s$^{-1}$. The previously derived value of $v
\sin i$ is smaller than the velocity resolution of our data. We
rotationally broaden the synthetic spectra by $v \sin i = 15$ km
s$^{-1}$, which is the convolution of the true $v \sin i$ ($\sim 5$ km
s$^{-1}$) and the instrumental resolution ($\sim 14$ km
s$^{-1}$). Finally, we add continuum veiling to these rotationally
broadened synthetic spectra and find the veiling value that provides
the best match between observed and synthetic data.  Fitted veilings
are $0.61 \pm 0.09$ and $0.57 \pm 0.04$ in January and May,
respectively.  The two measurements are thus consistent within the 
1-$\sigma$ statistical errors.

\begin{figure}[tbhp]
\plotone{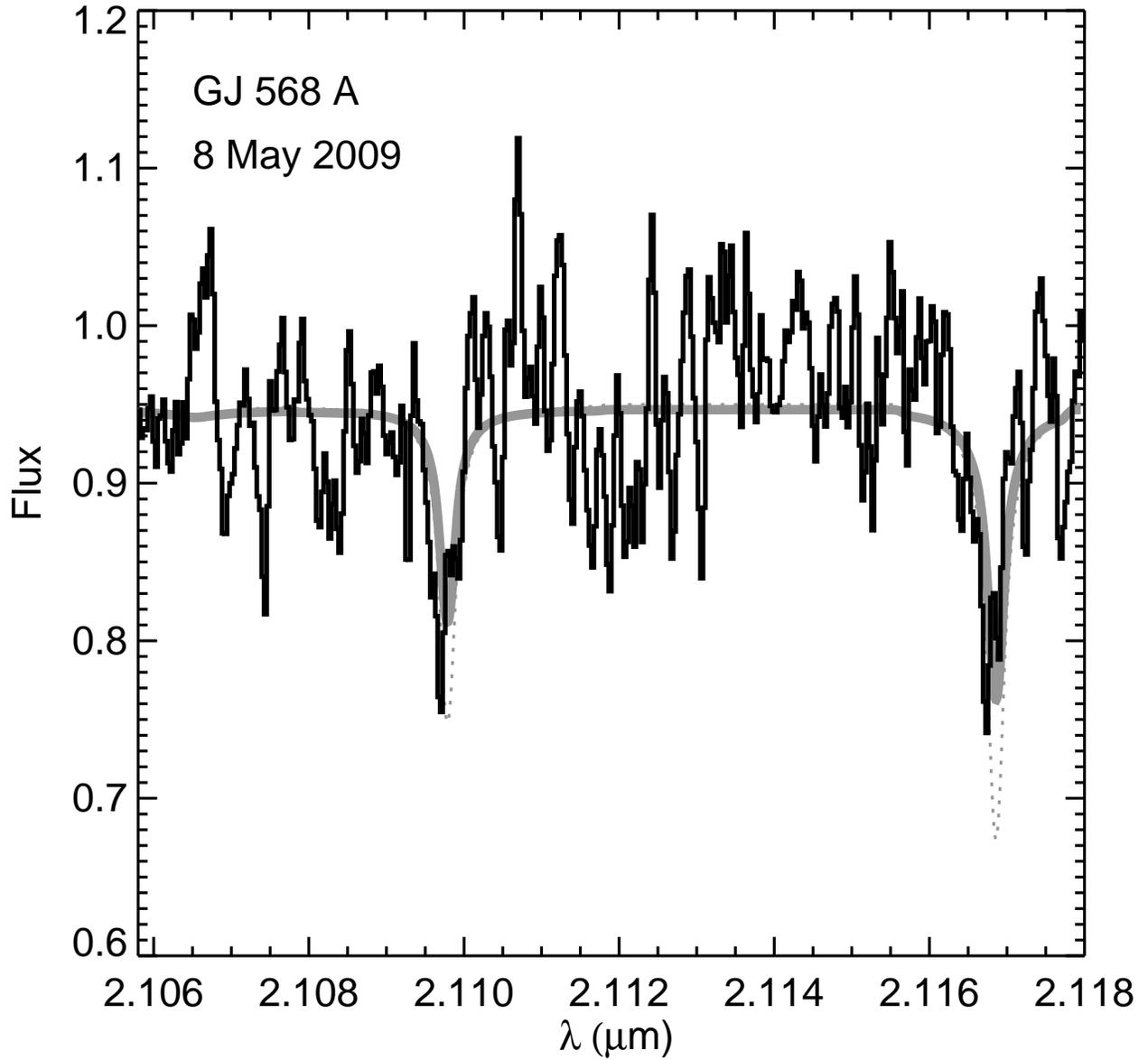}
\caption{Spectrum of GJ 568 A (solid black histogram) and a model spectrum that
  has been rotationally broadened and veiled (thick gray curve). We also plot the unveiled
  synthetic spectrum with a gray dotted curve.
\label{fig:gj568veiling}}
\end{figure}

\citet{DOPPMANN+05} used this procedure to derive veilings from NIRSPEC
data.  They tested the method with observations of unveiled MK standard stars,
and found that a correction was needed to produce zero veilings for
these stars.  They suggested that scattered light inside the NIRSPEC
instrument may have led to an instrumental veiling contribution.
Since the ARIES spectrograph has not been characterized in this way,
we need to test for such systematic effects.
Using our observed spectrum of GJ 568, which is an unveiled
M3V main-sequence star \citep{GRAY+03}, we can constrain any possible
instrumental veiling.  We estimate the veiling for GJ 568 using the
procedure described above with $T_{\rm eff}=3500$ K, $\log g=5$, 
and $v \sin i = 15$ km s$^{-1}$  (Figure \ref{fig:gj568veiling}).
We find $r_K = 0.47 \pm 0.04$.

If we assume that the instrumental veiling is independent of spectral
type, at least across the M3 to K7 range, then we can use the derived
veiling for GJ 568 A to correct the veilings inferred for TW Hya.  The
corrected veilings for TW Hya are $0.14 \pm 0.10$ and $0.10 \pm 0.06$
in January and May, respectively. The inferred veilings in the two
epochs are consistent with each other and with
with a previous measurement by \citet{JV01}.

\section{Discussion \label{sec:disc}}
\subsection{Inner Disk Variability}
The variation in Br$\gamma$ emission line-strength between our two observed
epochs, and between previous measurements in the literature, suggests that
the rate at which gas is being accreted by the central star is changing with time.
However, the amount of dusty material in the inner disk does not
appear to change significantly, based on the inferred veilings.
Our measured photometry is consistent with the 2MASS
magnitude, providing further evidence that the amount of matter in the
inner disk of TW Hya is not changing substantially with time.

\subsection{Disk Clearing Mechanisms \label{sec:clearing}}
Several mechanisms for creating the optically thin clearing in the TW
Hya disk have been proffered \citep[e.g.,][]{NAJITA+10}, including planets,
which may clear gaps about their orbits
\citep{CALVET+02}; photoevaporation of inner
disk material \citep[e.g.,][]{ACP06,OWEN+10}; or grain growth,
which would deplete the population of small grains that would produce
the near-IR emission.  The accretion rates measured here are difficult to
reconcile with EUV-driven photoevaporation scenarios \citep{ACP06},
although not necessarily with EUV+X-ray-driven photoevaporation models
\citep{OWEN+10}.  The sharp edge of the TW Hya disk
\citep[e.g.,][]{CALVET+02,HUGHES+07} seems to argue against grain
growth.  

While a planetary explanation may be favored, it too is
problematic since the massive outer disk of TW Hya should cause rapid
migration and destruction of planet-mass objects unless the disk is
unusually inviscid \citep[e.g.,][]{ECH06}.  While a 10
Jupiter-mass planet has been claimed to orbit TW Hya at 0.04 AU
\citep{SETIAWAN+08}, this would not dynamically effect the outer disk
edge at $\sim 4$ AU.  Moreover, follow-up observations demonstrate
that the claimed planetary signal is more likely due to starspot noise
\citep{HUELAMO+08}.

Any of the mechanisms discussed above are compatible with
time-variable accretion.  Time-variable accretion in the outer
disk--for example, due to changes in MRI turbulence--can lead to a
larger flow of material through the inner disk, even if
the flow is impeded by a planet or photoevaporation.
In fact, the inner disk accretion rate can vary
indpendently since it too may be driven by MRI turbulence
\citep{CM07}.

%Once a hole is created--by any mechanism--there are viable theories
%for how to maintain the hole.  One isThe other mechanism is the
%magnetorotational instability (MRI).  The inner edge 
%of the optically thick disk presents 
%a large surface area to stellar X-rays, allowing a high enough ionization
%for the MRI to be activated, and leading to the rapid accretion of 
%material from the inner disk edge through the cleared region and
%onto the star \citep{CM07}.    

While our data indicate a changing accretion rate with time, the low
veilings observed at both epochs indicate a small and nearly constant
amount of dust in the optically thin region.  A constant dust content
with time is also compatible with the unchanging $K$-band magnitude
inferred from our data.  

In fact, the lack of optically thick dust in the inner disk--at any
observed epoch--is surprising if one makes a simple estimate of the
disk column density.  For this calculation, we assume $T_{\rm disk}(R)
\propto R^{-1/2}$ and $\Sigma_{\rm disk}(R) \propto R^{-3/2}$.  We
take the disk scale height to be a constant fraction of the radius,
$H_{\rm disk} = 0.1 R$.  The pressure in the disk is thus 
$P_{\rm  disk}(R) \propto \rho_{\rm disk} T_{\rm disk} \propto R^{-3}$.

For the accretion rates derived here ($\sim 10^{-9}$ M$_{\odot}$
yr$^{-1}$), and assuming a viscous disk with $\alpha=0.01$, we derive
a disk surface density at $R= 1$ AU of 
\begin{equation}
\Sigma_{\rm disk} = \left(\frac{\dot{M}}{3\pi \alpha}\right)
\left(\frac{\mu m_{\rm H}}{k T_{\rm disk}}\right) \frac{v_{\rm
    Kepler}}{R}
\approx 5 {\rm \: g \: cm^{-2}},
\end{equation}
and a disk density of
\begin{equation}
\rho_{\rm disk} = \frac{\Sigma_{\rm disk}}{0.1 R} \approx 3 \times
10^{-12} {\rm \: g \: cm^{-3}}.
\end{equation}
If we assume $\kappa_{\rm dust} = 10^3$ \citep[appropriate for
sub-micron-sized dust; see, e.g.,][]{MN93},
and take a gas-to-dust ratio of 100,
then the vertical optical depth of the disk at 1 AU is
\begin{equation}
\tau \sim \left(\frac{\rho_{\rm disk}}{100}\right) H_{\rm disk}
\kappa_{\rm dust} \approx 15.
\end{equation}
Even for accretion rates lower by an order of magnitude, such as those
inferred in previous studied (Section \ref{sec:intro}), $\tau \ga 1$ for
this simple, viscous disk calculation.

To explain the lack of dust in the inner disk, we propose three
possible scenarios, all of which may operate in the TW Hya system.
The first is that the density could be over-estimated.  
An $\alpha$--disk may not be the correct description
for the inner disk of TW Hya, or a higher value of $\alpha$ may be
appropriate.  A  gas-to-dust ratio $>100$ could also help to maintain
a higher accretion rate with a smaller column of dusty material.
However, observations suggest that the gas-to-dust ratio in TW Hya--at
least as measured in the outer disk--may be substantially {\it smaller} than 100
\citep{THI+10}. 
 
The second possible explanation is dust-filtering, where the
pressure gradient at the inner edge of the optically-thick disk leads
to super-Keplerian velocities, and hence trapping, of dust
particles in a certain particle-size range \citep[e.g.,][]{RICE+06}.  
By preventing dust from reaching the inner disk, filtering can
maintain an optically thin region.  Moreover, even if the gas flow
through the inner disk varies, filtering may trap (some of) the
additional dust particles in the flow, preventing variations in
observed veiling.

Finally, radiation pressure from the star can help to drive dust
particles out of the inner disk.  Alone, this mechanism is unlikely to
clear out the inner disk or TW Hya, since an optically thick disk
can be replenished by accretion faster than material can be blown out
\citep[e.g.,][]{TL03}.  However, if the inner disk is optically thin
to begin with, for example due to dust filtering, then radiation
pressure acts directly on all dust in the system and can be an
efficient means of removal \citep[e.g.,][]{TA01,ECH06}.

Radiation pressure and dust filtering act in a complementary way.
Dust filtering is most efficient for particles with sizes of tens of
microns, since these are not too well coupled to the gas but don't
have the inertia of larger bodies \citep{RICE+06}. Radiation pressure
is most effective for smaller particles, with sizes $\la 1$ $\mu$m
\citep[e.g.,][]{WEIDENSCHILLING77}.  Thus, small particles that pass
through the ``filter'' into the inner disk may be pushed back out by
radiation pressure.  The
combination of filtering and radiation pressure may be sufficiently
efficient to maintain an inner disk virtually devoid of dust even
during periods of enhanced gaseous accretion.

\clearpage

\begin{deluxetable}{lcccccc}
\tabletypesize{\scriptsize}
\tablewidth{0pt}
\tablecaption{Derived Properties
\label{tab:results}}
\tablehead{\colhead{Epoch} & \colhead{$m_K$} & \colhead{Br$\gamma$ EW (\AA)} 
& \colhead{L$_{\rm Br\gamma}$ (L$_{\odot}$)} & \colhead{L$_{\rm acc}$ (L$_{\odot}$)} & 
\colhead{$\dot{M}$ (M$_{\odot}$ yr$^{-1}$)} & \colhead{$r_K$}}
\startdata
17 Jan 2009 & $7.28 \pm 0.02$ & $-7.5 \pm 0.9$ & $(3.4 \pm 0.5) \times
10^{-5}$ & $0.06 \pm 0.01$ &
$(3.0 \pm 0.5) \times 10^{-9}$  & $0.14 \pm 0.10$ \\
8 May 2009 & $7.29 \pm 0.05$ & $-3.6 \pm 0.4$ &  $(1.6 \pm 0.3) \times
10^{-5}$ & $0.03 \pm 0.01$ &
$(1.2 \pm 0.3) \times 10^{-9}$  & $0.10 \pm 0.06$ \\
\enddata
%\tablecomments{While the veilings inferred from the two epochs are
 % consistent with one another, they may be biased by systematic
 % effects (as discussed in \S \ref{sec:veiling}).  We therefore are
 % not confident that the absolute values of veilings are correct, or
 % necessarily inconsistent with zero.
%}
\end{deluxetable}

\epsscale{1.0}

%\epsscale{1.0}
%\begin{figure}[tbhp]
%\plotone{figs/twhya_veiling_compare}
%\caption{Spectra of TW Hya observed over two epochs (solid and dashed
 % black histograms) and the spectrum of GJ 658 A observed on 8 May
 % 2009 (thick gray histogram).  All three spectra appear fairly
 % similar to each other, although there are some small differences due
  %to the different spectral type and airmass of the two sources, and
  %noise in the data.
  %Given that GJ 568 A is an unveiled main-sequence star, this
  %comparison suggests veilings close to zero in both epochs of data
  %for TW Hya.
%\label{fig:kveiling_comp}}
%\end{figure}

\epsscale{1.0}
\begin{figure}[tbhp]
\plottwo{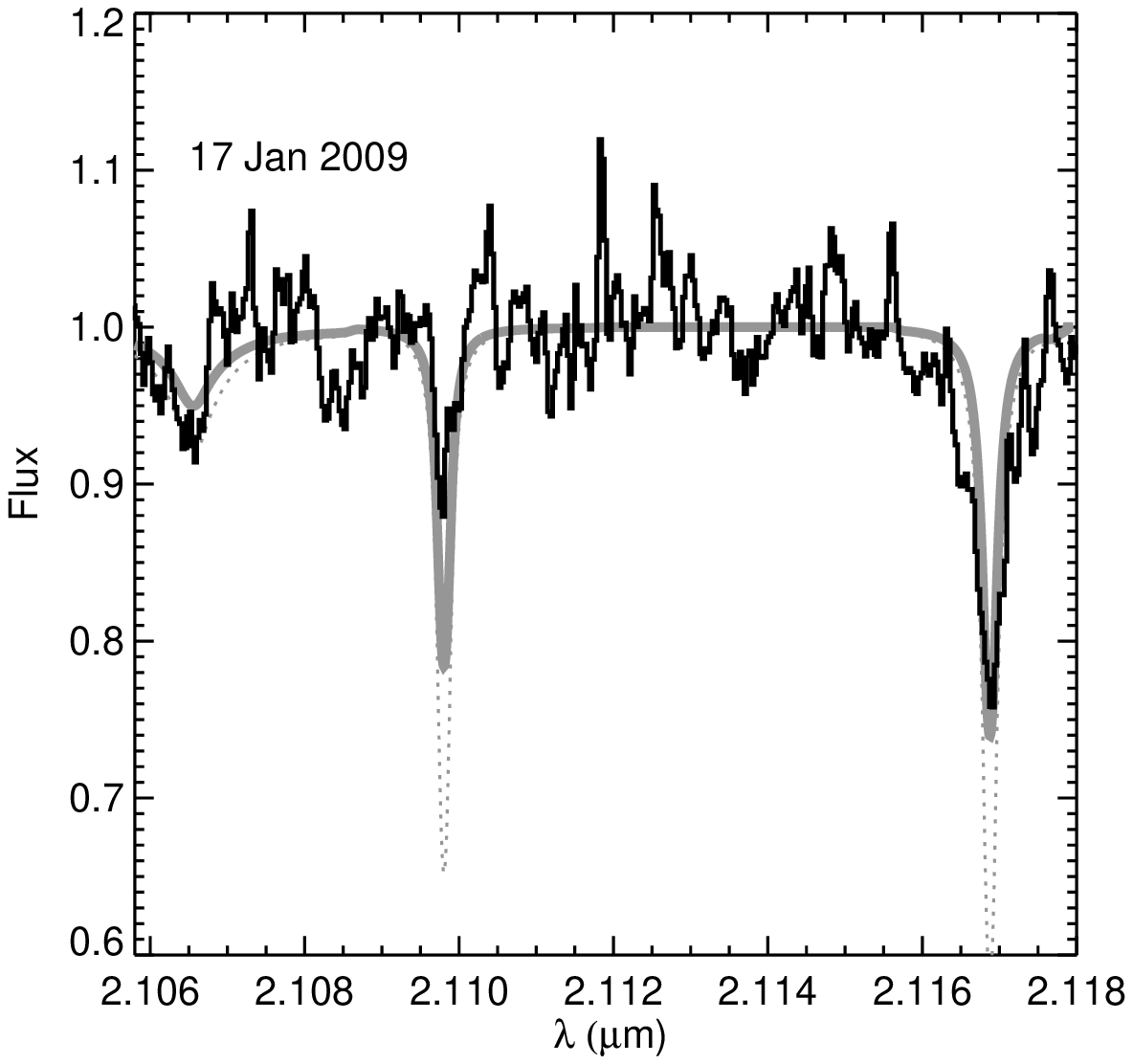}{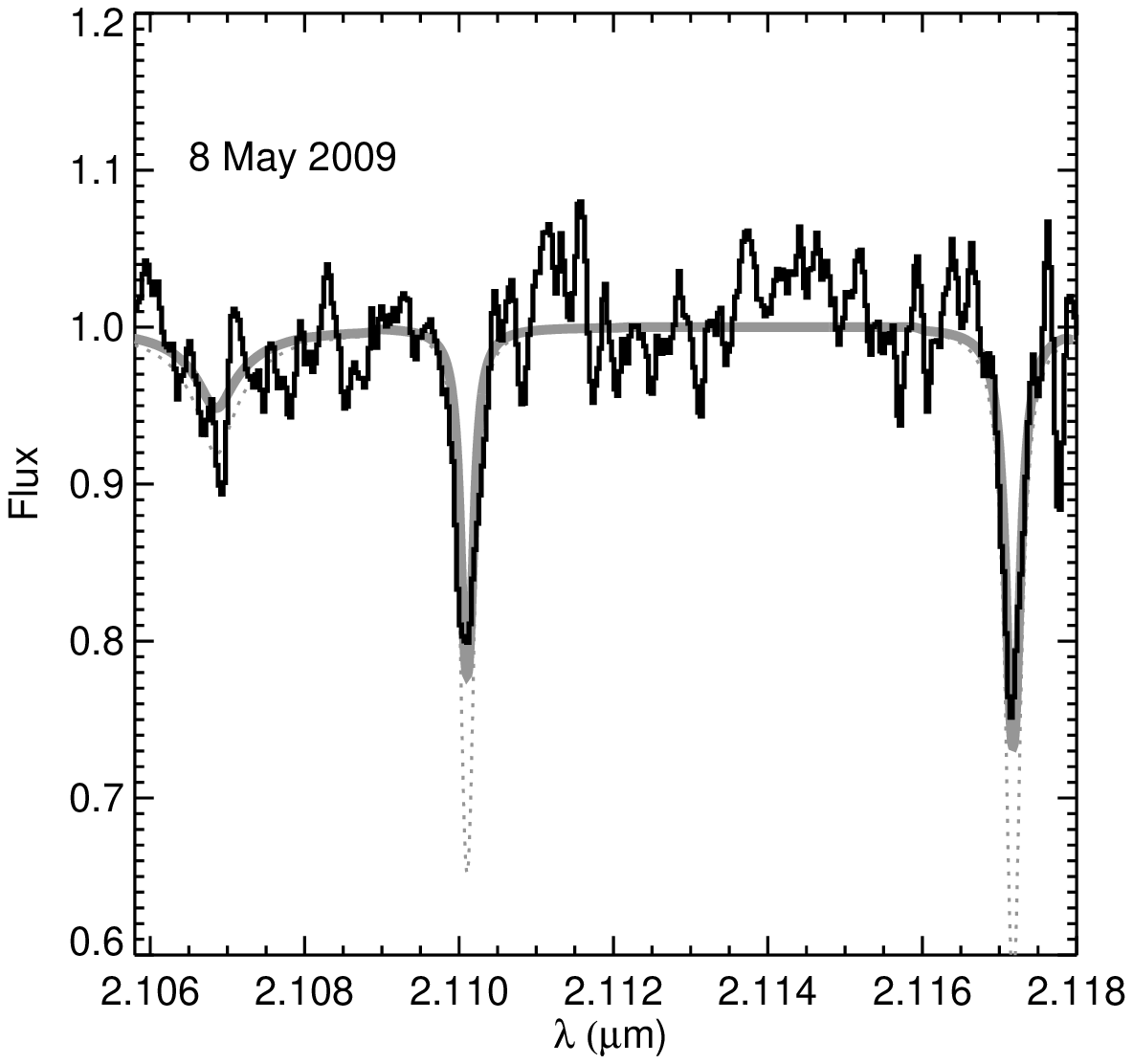}
\caption{Spectra of TW Hya (solid black histograms) and model spectra that
  have been rotationally broadened and veiled (thick gray curves), for
  17 January (left) and 8 May (right).   We also plot the unveiled
  synthetic spectra with gray dotted curves.  As discussed in \S
  \ref{sec:veiling}, systematic errors, in particular instrumental scattering, 
  lead to inferred veilings larger than true values.
\label{fig:kveiling}}
\end{figure}

\epsscale{1.0}

\end{document}